\pdfoutput=1

\documentclass[aps,prl,preprint,groupedaddress]{revtex4}
\usepackage{amsmath}
\usepackage{graphicx}

\usepackage{color}

\usepackage[colorlinks=true,
            urlcolor=blue,
            citecolor=blue,
            linkcolor=black,
            pdfstartview=FitH,
            pdfpagemode=None]{hyperref}

\def\bra#1{\mathinner{\langle{#1}|}}
\def\ket#1{\mathinner{|{#1}\rangle}}
\newcommand{\braket}[2]{\langle #1|#2\rangle}
\newcommand{\ketbra}[2]{|#1\rangle\langle#2|}

\newcommand{\mbf}[1]{\mathbf{#1}}
\newcommand{\up}{\uparrow}
\newcommand{\dn}{\downarrow}

\begin{document}


\title{Coherent control of atomic transport in spinor optical lattices}


\author{Brian Mischuck} \email{bmischuc@unm.edu}
\affiliation{University of New Mexico}
\author{Poul S. Jessen} 
\affiliation{University of Arizona}
\author{Ivan H. Deutsch}
\affiliation{University of New Mexico}


\date{\today}

\begin{abstract}
Coherent transport of atoms trapped in an optical lattice can be controlled by microwave-induced spin flips that correlate with site-to-site hopping.  We study the controllability of homogeneous one-dimensional systems of noninteracting atoms in the absence of site addressability.  Given these restrictions, we construct a deterministic protocol to map an initially localized Wannier state to a wave packet that that is coherently distributed over $n$ sites.  This is extended to analytic solutions for arbitrary unitary maps given homogenous systems and in the presence of time-dependent uniform forces.  Such control is important for applications in quantum information processing such as quantum computing and quantum simulations of condensed matter phenomena.
\end{abstract}


\maketitle

\section{I. Introduction}
Neutral atoms trapped in optical lattices have emerged as a rich platform for exploring a wide variety of phenomena and devices based on coherent quantum dynamics.  Examples include quantum computers \cite{brennen_quantum_1999, jaksch_entanglement_1999, mandel_controlled_2003, nelson_imaging_2007}, quantum simulators of condensed matter \cite{jaksch_cold_2005, lewenstein_ultracold_2007}, topological quantum field theory \cite{duan_controlling_2003, micheli_toolbox_2006}, and quantum chaotic dynamics \cite{ steck_observation_2001, ghose_atomic_2001, argonov_nonlinear_2008}.  An essential ingredient in these systems is the coherent control of atomic transport in the lattice.   Such transport is driven by time-dependent variations in the lattice potential and the application of external fields. In its most basic form, the atoms' ballistic tunneling between sites in a sinusoidal potential can be controlled through time-dependent modulations of the lattice depth and phase.  The latter can be used to impart a time-dependent acceleration to the lattice, thereby simulating the effects of an applied electric field for electrons in a crystal that give rise to the fundamental paradigms of coherent transport in solid-state physics.  Bloch oscillations \cite{ben_dahan_bloch_1996}, Wannier-Stark ladders \cite{wilkinson_observation_1996}, Landau-Zener tunneling \cite{zenesini_time-resolved_2009}, and dynamical localization \cite{Madison_1998} have all been demonstrated in optical lattices and explored as mechanisms for coherent control.

More complex lattice geometries introduce additional features.  For example, in a lattice of double-wells, one can drive transport between sites in a pairwise manner, assuming a sufficient barrier to ignore tunneling between different double-wells \cite{sebby-strabley_lattice_2006, folling_direct_2007, romero-isart_quantum_2007,de-chiara08a}.  In this case, the control problem is substantially simplified, as the relevant Hilbert space in a given time interval is restricted to a small discrete set of energy levels, as opposed to the infinite chain of levels in a sinusoidal lattice. Control across the entire lattice can be implemented by modifying the geometry so that the wells are alternatively coupled to all nearest neighbors (left or right in 1D).  Designer double-well lattices have been explored for quantum information processing tasks such as quantum computing \cite{anderlini_controlled_2007} and simulations of condensed-matter phenomena \cite{bloch_quantum_2008-1}.  

Still richer control is possible for spinor lattices where the optical potential depends on the atom's internal spin state \cite{haycock_mesoscopic_2000}.  The lattice's morphology can now be modified through variation of the laser polarization as well as intensity, lattice phase, etc.  The earliest proposals for quantum logic in optical lattices via controlled collisions involved transport of the atoms via time-dependent rotation of the direction of a laser beam's polarization \cite{brennen_quantum_1999, jaksch_entanglement_1999,mandel_coherent_2003}.  An alternative and perhaps more robust route to coherent control of atomic transport is to use external fields to drive spin-changing transitions that are correlated with atomic motion, similar to the scheme proposed by Foot {\em et al.} \cite{deb_method_2007}.  Such protocols can make use of the tools for robust control of spins \cite{vandersypen_nmr_2005, li_control_2006,rakreungdet_accurate_2009}, as developed in NMR, to the control of atomic motion in the lattice.

In this article we explore methods for coherent control of atomic transport with microwave-induced spin rotations between hyperfine levels and polarization-gradient lattices.  Our main focus is on controllability -- how the Hamiltonian that governs the dynamics restricts the possible unitary maps that one can implement, and how to design specific waveforms to carry out a given task.  We will consider here the simplest problem of noninteracting atoms in one dimension.  While extensions to the interacting case are nontrivial, the current work is an important stepping-stone in that direction.  

The remainder of the paper is organized as follows.  In Sec. II we establish the formalism necessary to describe spinor lattices and their interaction with external fields.  We apply this to study the conditions for wave function control (the preparation of a desired spinor wave function starting from a known localized Wannier state) and prescribe a constructive algorithm for carrying out this task in Sec. III.  We then generalize this in Sec. IV to the case of more general unitary maps for unknown initial states.  Finally, we summarize and give an outlook towards future research in this area in Sec. V.

\section{II. Microwave-Driven Spinor Lattices}
Spinor lattices arise from the tensor nature of atom-photon interaction.  In a monochromatic laser field $\text{Re}\left( \mbf{E}(\mbf{x}) e^{i \left( \phi(\mbf{x})- \omega_L t \right)} \right)$, the light-shift potential takes the form,
\begin{equation}
V_{LS}(\mbf{x})= -\frac{1}{4} \alpha_{ij} E^*_{i}(\mbf{x}) E_{j}(\mbf{x}),
\end{equation}
where $\alpha_{ij}$ is the atomic dynamic polarizability at frequency $\omega_L$, for atoms in a particular ground-state manifold.  We consider a 1D geometry consisting of counterpropagating laser beams with linear polarizations, forming a relative angle $\theta$ (the ``lin-$\theta$-lin" geometry).  Restricting our attention to alkali atoms, when the laser field detuning is large compared to the  excited-state hyperfine splitting but not large compared to the fine-structure splitting, ellipticity in the laser field leads to a fictitious magnetic field that varies periodically in space, and results in a spin-dependent light shift \cite{deutsch_quantum-state_1998}.   Taking the quantization axis along one of the laser beams' wave vectors, the spinor light-shift potential can then be written
\begin{equation}
V_{LS} = \sum_{F,m_F} \left[ E_F -V_{F,m_F} \cos \left( 2k_L z + \delta_{F,m_F} \right) \right] \ket{F, m_F}\bra{F, m_F},
\end{equation}
where $E_F$ is the degenerate energy of a hyperfine manifold, $V_{F,m_F} = V_0 \sqrt{\cos^2 \theta + \left(g_F m_F/2 \right)^2 \sin^2 \theta }$,  and $\delta_{F,m_F} = \tan^{-1} \left( g_F m_F \tan\theta / 2 \right)$.  Here, $V_0$ is the lattice depth for lin-$\parallel$-lin and $g_F$ is the Land\'{e} g-factor of the hyperfine manifold under consideration.

The addition of a static bias magnetic field breaks the degeneracy between Zeeman sublevels within a manifold and allows us to spectrally isolate different microwave transitions between the manifolds.  For concreteness, we consider $^{133}$Cs,  and choose a two-state subspace $\ket{F=4,m=3} \equiv \ket{\up}, \ket{F=3,m=3} \equiv \ket{\dn}$ to define a pseudo-spin-1/2 particle.  Restricting to this subspace, and adding a near-resonant microwave field with magnetic field $B_{\mu w} \cos \left( \omega_{\mu w} t - \phi \right)$ that couples these spin states, the Hamiltonian in the rotating frame takes the form $H  =  H_{latt} + H_{\mu w}$, where
\begin{subequations}
\begin{align}
H_{latt}& = \frac{p^2}{2m}  -  V_{0}\cos \left( 2k_Lz+\delta_{0} \right)\ketbra{\up}{\up} - V_{0}\cos \left( 2k_L z-\delta_{0} \right)\ketbra{\dn}{\dn}\\
H_{\mu w} &=  -\frac{\Delta_{\mu w}}{2}\sigma_z -\frac{\Omega_{\mu w}}{2}(\cos\phi_{\mu w} \,  \sigma_x + \sin\phi_{\mu w} \, \sigma_y).
\end{align}
\label{eq:hamiltonians}
\end{subequations}
Here $V_0 =V_{4,3} \approx V_{3,3}$, $\delta_0=\delta_{4,3} \approx -\delta_{3,3}$, $\Omega_{\mu w} = \bra{\up} \hat{\mu} \ket{\dn} B_{\mu w} / \hbar$ is the microwave resonant Rabi frequency, $\Delta_{\mu w}$ is the microwave detuning from a hyperfine resonance defined by the untrapped atoms, $\phi_{\mu w}$ is the phase of the microwave oscillator, and the Pauli-$\sigma$ operators are defined relative to the pseudo-spin.  

Neglecting the kinetic energy and diagonalizing the Hamiltonian leads to adiabatic or microwave-dressed potentials ,
\begin{equation}
V_\pm(z) = -V_0 \cos\delta_0 \cos(2kz) \pm \frac{1}{2}  \sqrt{\left(2V_0 \sin\delta_0 \sin(2kz) - \Delta_{\mu w} \right)^2 + \Omega_{\mu w}^2}.
\end{equation}
At $\Delta_{\mu w}=0$, in the lin$\perp$lin ($\theta= \pi/2$) geometry, the adiabatic potentials yield a period $\lambda/4$ ``subwavelength" lattice \cite{lundblad_atoms_2008,yi_state-dependent_2008,shotter_enhancement_2008}.  In the context of a Hubbard Hamiltonian describing interacting particles moving on a lattice \cite{jaksch_cold_1998}, this configuration gives us greater freedom to independently control the site-to-site tunneling rate $J$ and the onsite interaction strength $U$ \cite{zenesini_coherent_2008}. By employing both optical and microwave fields, the lattice depth dominates the control of  $U$ while the the applied microwave dominates control of $J$. Moreover, the tunneling matrix element is complex, set by the microwave phase, allowing for time-reversible tunneling and further control \cite{cucchietti_loschmidt_2006}.  
 
For $\theta \ne n\pi/2$, the adiabatic potentials take the form of a lattice of double-well potentials arising from the asymmetry for transport to the left vs. the right. The parameters characterizing the double well, including barrier height, tunneling matrix element, and energy asymmetry (``tilt"), can be controlled through variations of lattice intensity/polarization, microwave power, and detuning.  The richness of this system should enable us to control wave function coherence for spinors over multiple sites.  Our early work on this subject demonstrated spinor double-well coherence driven by Larmor precision in a quasistatic magnetic field \cite{haycock_mesoscopic_2000}.  The current approach, based on applied microwave fields should be much more robust and controllable.

While the dressed-lattice adiabatic potentials guide intuition about the transport, quantitative predictions are more accurately made by considering the band structure of the Hamiltonian, $H_{latt}$, in Eq. (\ref{eq:hamiltonians}).  Associated with the spin $s=\up$ and $s=\dn$ lattices are Bloch states for band-$n$ and quasimomentum-$q$, $\ket{\psi^{(s)}_{n,q}}$, and Wannier states for that band and lattice site-$l$,  $\ket{\phi^{(s)}_{n,l}}$, related by the usual Fourier transform over the first Brillouin zone,
\begin{equation}
\ket{\phi^{(s)}_{n,l}} =  \int_{-1/2}^{1/2}e^{i 2 \pi l q} \ket{\psi^{(s)}_{n,q}} dq.  
\end{equation}
Here and throughout, lengths are measured in units of the lattice period $L=\lambda_L/2$ and wave numbers in units of the reciprocal lattice vector $K=4 \pi/ \lambda_L$.  For sufficiently deep lattices and atoms in the lowest lying bands, tunneling between sites is completely negligible over the timescales of interest.  In that case, the lattice Hamiltonian is diagonal both in the Bloch and Wannier bases, with no energy variation over the $q$ or $l$ index.  

\begin{figure}
\includegraphics{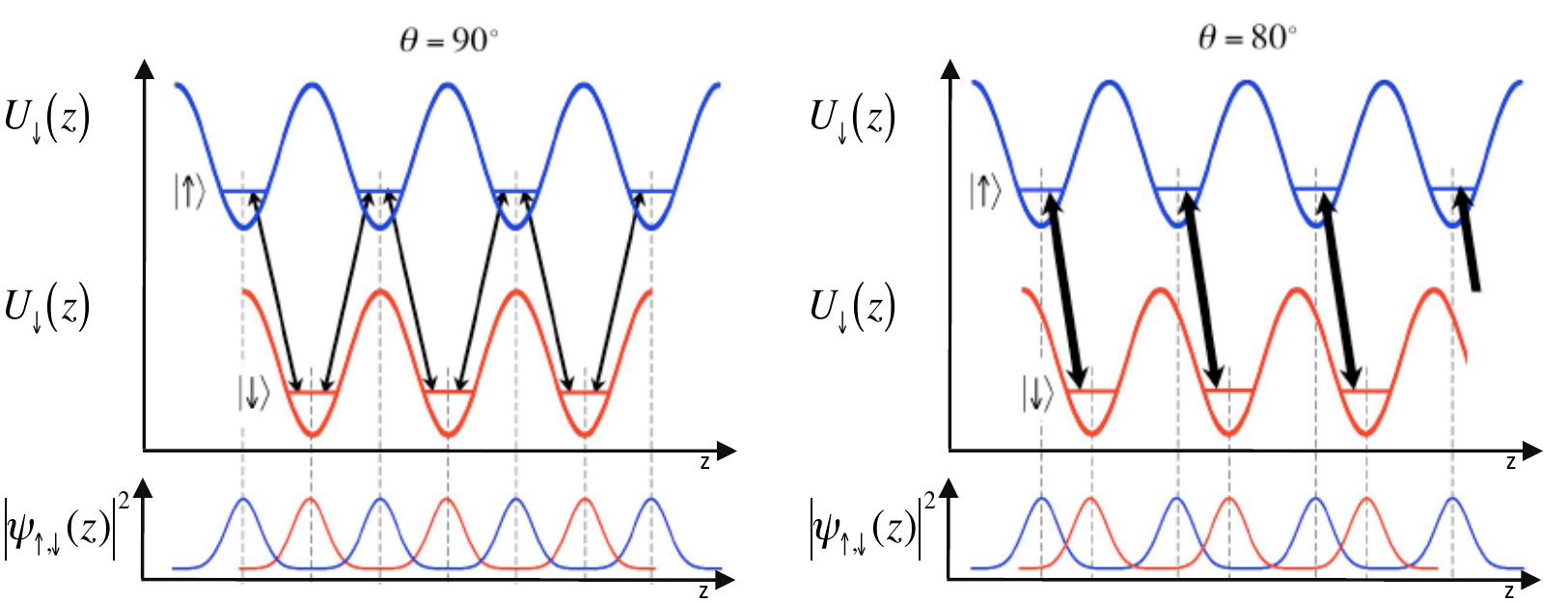}
\caption{\label{fig:ShiftedLattices2} Controlled transport in lin-$\theta-$lin.   Spin-dependent lattices have a relative phase shift due to polarization gradient.  In the lin$\perp$lin configuration, $\theta=90^\circ$,  there is no asymmetry in right-left transport, and atoms can ballistically tunnel in the dressed potential. For a change of just $10^\circ$ degrees away from lin$\perp$lin, in a lattice with an oscillation frequency of 20$E_R$, the ratio of the effective tunelling rates is  $\Omega_L /  \Omega_R \approx 3500$.  The result is lattice of double wells with pair-wise tunneling couplings.}
\end{figure}

Transport dynamics are driven by the microwaves tuned to cause transitions between the ground bands associated with the spin-up and spin-down lattices.  We assume that the detuning and Rabi frequency are sufficiently small so that single-band/lattice tight-binding (TB) model is a good approximation.  Henceforth we drop the band index and set $n=0$.   In the Wannier basis, the total Hamiltonian in the TB approximation is
\begin{equation}
H_{TB} = \sum_{l=-\infty}^\infty  -\frac{\Delta_{\mu w}}{2} \sigma^l_z -\frac{1}{2} \left[ e^{i \phi_{\mu w}} \left(\Omega_R \sigma^{l,R}_{+} + \Omega_L \sigma^{l,L}_{+} \right) +h.c. \right], 
\label{eq:H_TB}
\end{equation}
where, 
\begin{subequations}
\begin{align}
\sigma_z^l &\equiv \ket{l,\up}\bra{l,\up} - \ket{l,\dn}\bra{l,\dn},\\
\sigma_+^{l,R} &\equiv \ket{l,\up}\bra{l,\dn},\\
\sigma_+^{l,L} &\equiv \ket{l-1,\up}\bra{l,\dn}
\end{align}
\end{subequations}
are the Pauli operators for two-level transitions that pairwise couple spin-down Wannier states to their neighbors on the right, $\ket{l,\dn} \rightarrow \ket{l,\up}$, and on the left $\ket{l,\dn} \rightarrow \ket{l-1,\up}$.  Note, we have chosen an arbitrary labeling of the Wannier state indices by convention so that a spin-down state and spin-up state to its right are both associated with the same lattice period label, $l$.  Because the microwaves transfer negligible momentum to the atoms, translation of the atomic wavepacket is possible only when the probability amplitude of an atom overlaps between neighboring sites.  The Rabi frequencies for transitions to the left or right are thus weighted by Franck-Condon factors, $\Omega_R = \braket{\phi^{\up}_l}{\phi^{\dn}_l}\Omega_{\mu w}$, $\Omega_L = \braket{\phi^{\up}_{l-1}}{\phi^{\dn}_l}\Omega_{\mu w}$. For the ground bands in the TB approximation, a large asymmetry in right-left transport and isolation of double wells arises from small asymmetry in right-left displacement of the lattice due to the Gaussian overlap of the wavepackets (see Fig. \ref{fig:ShiftedLattices2}). 

The combination of spinor optical lattices and microwave-driven spin dynamics provides a wide variety of parameters that can be modulated in real time during an experiment to coherently control atomic transport.  In the next section we study the formal controllability of this system and develop constructive protocols to implement desired unitary maps.

\section{III. Wavepacket control}
Given the time-dependent Hamiltonian at hand, a first question to address is ``controllability", i.e., which class of unitary transformations can be generated by the arbitrary design of the waveforms that parameterize that Hamiltonian.  We consider first the problem of preparing an arbitrary wavepacket that is coherently distributed over multiple sites of the lattice, starting from an initially localized Wannier state.  Unless specially designed to allow for individual site addressability \cite{nelson_imaging_2007, zhang_manipulation_2006}, control in a typical lattice is limited by translational invariance of the operations.  Although a linear gradient breaks the translational symmetry,  the controllability of the system is still limited, as we show below. We thus restrict our attention to strictly periodic lattices and allow for an additional constant force $F$ on the atoms, as in the case of lattices held vertically in gravity, or when the overall lattice is accelerated through time-dependent changes of the standing wave pattern.

We consider Hamiltonians which are composed of a translationally invariant part, $H_0$, with period $L$, and an applied force, $F$,
\begin{equation}
H(t) = H_0(t)+F(t)x
\label{eq:general_hamiltonian2}
\end{equation}
A particular example is  $H_0(t)=H_{TB}(t)$, as given in Eq. (\ref{eq:H_TB}), with time-dependent variations in microwave power and/or phase.  The time evolution of such a system may be written in the form of the time-ordered exponential 
\begin{equation}
U(t) = T \exp \left\{ -i \int_0^t \, \left( H_0(t')+F(t')x \right) dt' \right\},
\label{eq:unitaries}
\end{equation}
Such a map has the property that if we translate the entire system by $j$ times the period of $H_0$, then $U \rightarrow e^{-i\int_0^t F(t')\,dt' jL}U$.  If the initial state $\ket{\phi}$ is a localized Wannier state, it satisfies
\begin{equation}
\langle\phi| T_j|\phi\rangle = \delta_{j,0},
\label{eq:initial_state} 
\end{equation}
where $T_j$ translates the system by $jL$.  If state $\ket{\phi}$ maps to $\ket{\phi '}$ under a unitary evolution of this form, then the evolved state satisfies this same condition, as follows from the identity
\begin{equation}
\langle\phi'| T_j|\phi'\rangle = \langle\phi| U^{\dagger}T_j U | \phi\rangle = e^{i\int_0^t F(t')\,dt'jL} \langle\phi |T_j|  \phi\rangle= \delta_{j,0}.
\label{eq:reachable_states}
\end{equation}
Thus any wavepacket prepared by these controls must be orthogonal to itself after a translation by an arbitrary number of periods.  Furthermore, since the force $F$ has dropped out, the linear gradient does not impact the range of states that my be reached.

Are all states that satisfy this constraint reachable through some choice of the control waveforms that parameterize $H_{TB}$ in Eq. (\ref{eq:H_TB})?  To show that this is the case, we employ a protocol for constructing a desired state-to-state mapping as defined by Eberly and Law in the context of Jaynes-Cummings ladder \cite{law_arbitrary_1996}.  First note that if we can map the state $\ket{\phi}$ to another state $\ket{\phi'}$, and the control Hamiltonian allows the unitary map to be time-reversible, then we can map $\ket{\phi'}$ to $\ket{\phi}$.  Thus, in order to show that we can get from a localized state to any state satisfying Eq. (\ref{eq:reachable_states}), we consider the time-reversed problem of mapping such a state to the initial Wannier state.

\begin{figure}
\includegraphics{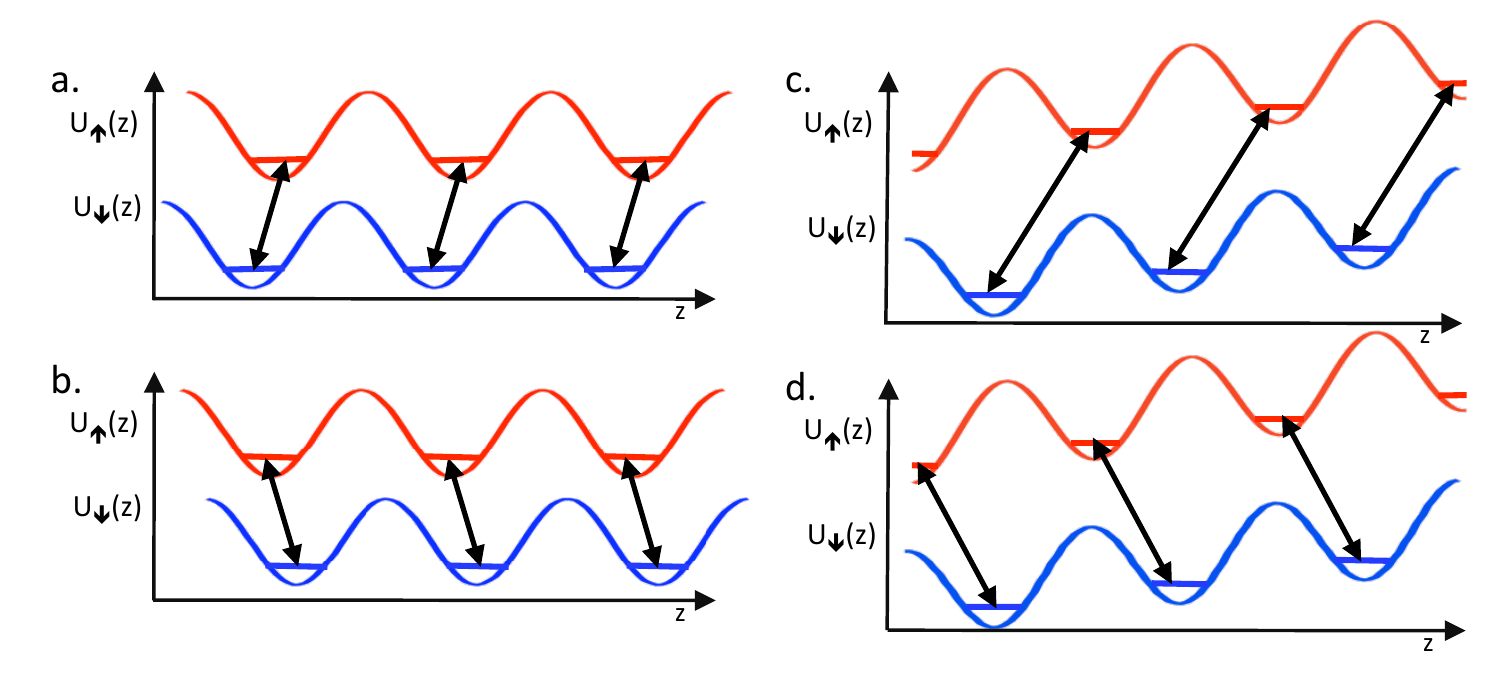}
\caption{\label{fig:TwoLevel} Two different methods for isolating unitary maps on two-level subspaces. In (a) and (b), the choice of polarization angles in a lin-$\theta$-lin lattice isolate different sets of two-level systems with transport either to the right or to the left.  In (c) and (d), two-level systems in a lattice with a linear gradient are spectrally addressed through their distinct microwave transition frequencies, which differ by $\pm FL/2\hbar$}
\end{figure}

To construct the desired map, we employ a series of $SU(2)$ rotations on resolvable subspaces of the total Hilbert space.  Such a collection of disjoint two-level systems can be addressed in a lin-$\theta$-lin spinor lattice in either an asymmetric configuration ($\theta \neq \pi/2$), or in a lin$\perp$lin configuration in the presence a sufficiently strong uniform force so that isolated pairs of states are spectroscopically addressable (see Fig. \ref{fig:TwoLevel}).   Note that in the latter case, because of the presence of an external force, if we translate the system by an integer multiple of $\lambda/2$, the state picks up an extra phase due to the linear gradient.  As we will see below, our construction does not require these phases, and since the construction is capable of synthesizing all reachable states, the phases are redundant.  We can ignore the phases if we choose the time over which the two level unitaries operate to be an integer multiple of $2\pi/FL,$ and we will assume this to be the case for the rest of this section.

We will restrict our attention to states with support strictly on a finite set of lattice sites and zero probability amplitude outside some range.  Such spinors can be represented as
\begin{equation}
\ket{\psi}=\displaystyle\sum_{l=l_{min}}^{l_{max}} \left( c_{l\downarrow}\ket{l\downarrow}+c_{l\uparrow}\ket{l\uparrow}\right) .
\label{eq:initial_state}
\end{equation}
The finite extent of the wave function, together with the condition expressed in Eq. (\ref{eq:reachable_states}), places a constraint on the two-level subspaces.  If we translate the entire state by $l_{max}-l_{min}$ then,
\begin{equation}
\bra{\psi}T_{l_{min}-l_{max}}\ket{\psi} = c_{l_{min}\uparrow}^* c_{l_{max}\uparrow}+c_{l_{min}\downarrow}^* c_{l_{max}\downarrow}=0.
\label{eq:outermost}
\end{equation}
Thus, in order to satisfy Eq. (\ref{eq:reachable_states}), the two outermost two-level subspaces must be orthogonal.  Moreover, because of translational invariance, the unitary transformations that we apply are equivalent at each period of the lattice. In particular, by unitarity, if we apply a rotation operator that maps the subspace on the left end of the atomic distribution to pure spin up, the subspace at the right end of the distribution must be rotated to pure spin down.   

\begin{figure}
\includegraphics{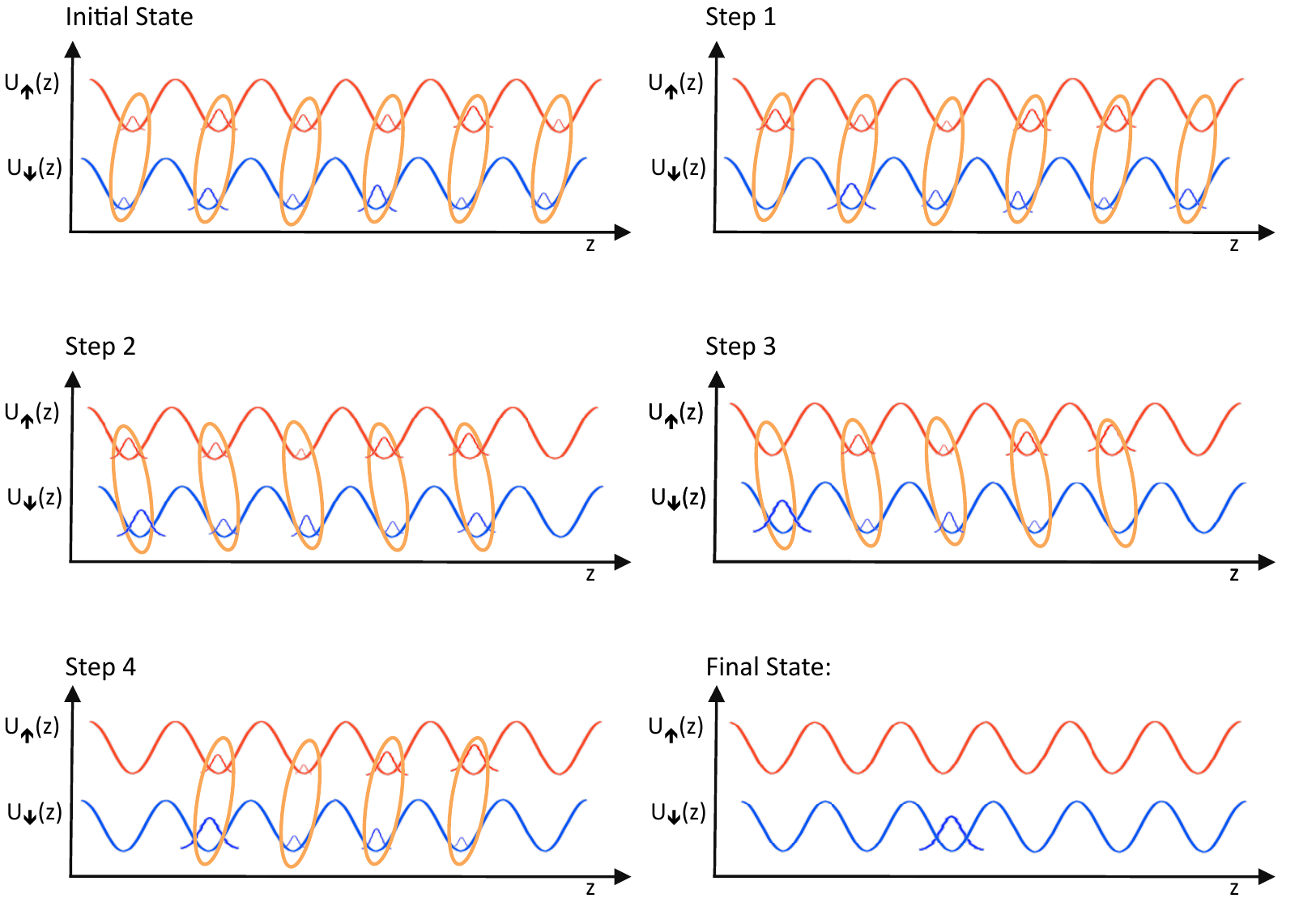}
\caption{\label{fig:StatePrep} An example of state preparation through sequential $SU(2)$ rotations.  The polarization of the two counterpropagating beams is chosen to isolate sets of two-level systems.  In step 1, population in the leftmost two-level system is mapped to entirely spin up, forcing population in the right most two level system to be entirely spin down by Eq. (\ref{eq:outermost}).  In step 2, the polarization is rotated so that a different set of two level systems are coupled.  In step 3, the population in the leftmost level is then mapped to entirely spin down.  In step 4 the polarization is set to its original configuration and steps 1-4 are repeated until the state is localized.  For this particular case, we have chosen to end the sequence spin down.  A different choice for the final pulse would have ended the state spin up.}
\end{figure}

These observations are the core of our construction (see Fig. \ref{fig:StatePrep}).   A sequence of two-level rotations can be used to map a coherent superposition delocalized across the lattice to one localized at a single site in a single spin state.  In the first step, a rotation is applied to map all population at the left-most two-level system ($l=l_{min}$) to spin-up according to the microwave-driven $SU(2)$ transformation,
\begin{equation}
S= \frac{1}{\sqrt{N}} \left[ \begin{array}{cc} c^*_{l_{min},\up} & c^*_{l_{min},\dn} \\ c_{l_{min},\dn}& -c_{l_{min},\up} \end{array} \right],
\end{equation}
where $N=\left| c_{l_{min},\up} \right|^2+\left| c_{l_{min},\dn} \right|^2$.  By the translation symmetry this simultaneously maps all population at the right-most two-level system ($l=l_{max}$) to spin-down, resulting in the spinor state
\begin{equation}
 \ket{\psi'}=c'_{l_{min}\up}\ket{l_{min}\up}+\displaystyle\sum_{l=l_{min}+1}^{l_{max}-1} \left( c'_{l\downarrow}\ket{l\downarrow}+c'_{l\dn}\ket{l\dn} \right) + c'_{l_{max}\dn}\ket{l_{max}\dn},
 \end{equation}
thereby shrinking the extent of the wave function by $\lambda/2$.  The lattice is then reconfigured (either through polarization rotation, a change of microwave frequency, or a change in acceleration) so that the opposite neighbors are coupled (spin-down now coupled to spin-up neighbor on left).  An appropriate $SU(2)$ rotation is then applied to map all population on the left most edge to spin down (simultaneously moving all population on right-most edge to spin-up).  Repeating, we form a sequence of rotations that take the outer edges of the distribution and map them inwards in steps of $\lambda/4$ until all the population is localized at one Wannier state.  Reversing the order of the sequence provides the desired protocol for constructing any single particle wave function in the given band and with support on a finite number of lattice sites, subject to the constraint  Eq. (\ref{eq:reachable_states}).  

\section{IV. Implementing general unitary transformations}
In the previous section we studied the construction of a particular unitary transformation --- mapping an initially localized Wannier state to a spinor wave function delocalized over a finite number of lattice sites, under the constraint of Eq. (\ref{eq:unitaries}). In this section we consider the most general unitary map that we can implement under these constraints.  We begin with the case of perfect translational invariance of the lattice.  In addition, since the microwave photons possess negligible momentum, only Bloch states with the same quasimomentum are coupled.  Because of these symmetries, any unitarity transformation that we can synthesize will be block diagonal in the Bloch basis, $U=\oplus_q U_q$, where each block is a $U(2)$ matrix that connects spin-up and spin-down states with the same quasimomentum $q$.  In the basis $\left\{ \ket{q,\up}, \ket{q,\dn} \right\}$,
\begin{equation}
U_q= e^{i\gamma(q)} \left[ \begin{array}{cc} \alpha(q) & \beta^*(q) \\ \beta(q) & -\alpha^*(q) \end{array} \right] ,
\label{generalU}
\end{equation}
where $|\alpha(q)|^2 + |\beta(q)|^2$=1. If the two lattices are sufficiently deep, tunneling is suppressed so $\gamma$ is independent of $q$ and can be factored out of the problem, leading to $SU(2)$ rotations in each block. 

The decomposition of a translationally invariant unitary transformation into a direct sum of $SU(2)$ matrices has important implications for the design of arbitrary maps.  Generally the design of a time-dependent waveform that generates an arbitrary unitary map is substantially more complex than a protocol for state-to-state mapping on an initially known state \cite{moore_relationship_2008}.  Intuitively, this is because state-to-state maps only constrain one column of a unitary matrix, whereas the evolution of the orthogonal complement is not fully specified.  The exception is for a spin-1/2 system.  By unitarity, specifying one column of an $SU(2)$ matrix necessarily constrains the other.  Since our spinor lattice is described by a collection of noninteracting spin-1/2 subspaces labeled by quasimomentum $q$, if we specify a state-to-state mapping of a spinor Bloch state, we specify the $SU(2)$ matrix on this block.  We can achieve this using the state-mapping protocol defined in Sec. III that takes an initially localized Wannier state to a state distributed over a finite number of lattice sites. Such a map specifies a transformation on each Bloch state according to the Fourier relationship between the probability amplitudes in the Wannier and Bloch bases. Based on this relationship, we can use our state-to-state map to design a more general class of unitary maps on the wave function.

To see this explicitly, consider the unitary evolution of an initial spin-up Wannier state (take $l = 0$ without loss of generality),
\begin{eqnarray}
U\ket{0,\dn} &=& \int_{-1/2}^{1/2} \, dq \left( \alpha(q) \ket{q,\up} + \beta(q) \ket{q, \dn} \right) \nonumber \\
&=&  \sum_{l=-\infty}^\infty \left( c_{l,\up}\ket{l ,\up} + c_{l,\dn}\ket{l,\dn} \right). 
\label{eq:l_synth_unitary}
\end{eqnarray}
The quasimomentum functions $\alpha(q)$ and $\beta(q)$ in Eq. (\ref{generalU}) are the Fourier sums of probability amplitudes in Wannier space, 
\begin{equation}
\alpha(q) = \sum_{l=-\infty}^{\infty} c_{l,\up} e^{-i2\pi lq},  \, \beta(q) = \sum_{l=-\infty}^{\infty} c_{l,\dn} e^{-i2\pi lq}. 
\label{eq:c_l_dn_syn}
\end{equation}
As long as the Fourier transform of $\alpha(q)$ and $\beta(q)$ have support only over a finite extent in $l$, we can generate these functions by applying the state-mapping protocol of the previous section to synthesize the probability amplitudes $c_l$ in Eq. (\ref{eq:c_l_dn_syn}).
For unitary maps defined by $\alpha(q)$ and $\beta(q)$ whose Fourier expansion in Wannier states does not have a strictly finite support, more general control methods are required.

We can easily generalize our result to include control through applied spatially uniform (possibly time-dependent) forces.  In the TB approximation, expressed in the Wannier basis, the Hamiltonian for a linear gradient potential in dimensionless units takes the form
\begin{equation}
H_{grad}(t) = \sum_{l=-\infty}^\infty F(t)L \left[ l\ketbra{l\downarrow}{l\downarrow}+(l+\delta l(t))\ketbra{l\uparrow}{l\uparrow}\right],
\end{equation}
where $\delta l(t)$ arises due to the off-set between spin-up and down lattices. We allow for modulations of the overall force as studied in ``shaken lattices" \cite{zenesini_coherent_2009,ivanov_coherent_2008,alberti_engineeringquantum_2008,eckardt_exploring_2009,sias_observation_2008}, and the possibility of time-dependent variations in the relative positions of the two spin states, as could be implemented through modulations in the laser beams' polarization direction.  The combination of this Hamiltonian, together with microwave-driven control described by $H_{TB}$ in Eq. (\ref{eq:H_TB}), gives rise to a general unitary transformation that can be written using the interaction picture in the form,
\begin{equation}
U(t) = D(t) U_I(t),
\label{eq:int_schrod}
\end{equation}
where,
\begin{equation}
D(t) = e^{-i\int_0^tH_{grad}(t')dt'} = e^{-i\chi/2}\int_{-1/2}^{1/2}\,dq\ketbra{q-\eta}{q}\otimes e^{-i\chi\sigma_z/2}
\end{equation}
with $\chi = \int_0^t\delta l(t') F(t') L\, dt'$,  $\eta=\int_0^t F(t') dt'$, and 
\begin{equation}
i\frac{d}{dt}U_I(t)  = H_I(t) U_I(t),
\end{equation}
where $H_I(t) = D^{\dagger}(t) H_{TB}(t) D(t)$.    The exact form of $H_I$ is rather complicated, but all that really matters for our argument is that it is translationally invariant with the period of the lattice, $T_j^{\dagger} H_I(t) T_j=H_I(t)$.  As a result, the solution, $ U_I(t)$ will be block diagonal in quasimomentum space, with the blocks consisting of $SU(2)$ rotations.   Thus by Eq. (\ref{eq:int_schrod}), the general unitary evolution will have the form,
\begin{equation}
U(t)  = e^{-i\chi/2} \int_{-1/2}^{1/2}\,dq\ketbra{q-\eta}{q}\otimes U_q.
\label{eq:unitary_lin}
\end{equation}
A control sequence of microwave driven rotations in a uniform lattice followed by a time dependent linear gradient can reach any unitary map of this form.

We contrast this control with that achievable in a one dimensional sinusoidal optical lattice with time-dependent uniform forces and modulation of the lattice depth in the TB approximation, without symmetry breaking for right vs. left transport.  Haroutyunyan and Nienhuis derived a general expression for the propagator in such a situation.\cite{haroutyunyan_coherent_2001}.  In the quasimomentum basis, the map is
\begin{equation}
U(t)\ket{q}=e^{-ia(t) \cos \left( 2 \pi q -b(t)\right)}\ket{q-\eta}
\end{equation}
where
\begin{equation}
a(t)e^{ib(t)} = \int_0^t \, dt' \, \Omega(t')e^{i\eta(t')}
\end{equation}
and $\Omega(t)$ is the hopping rate between sites (symmetric to the left or to the right), and $\eta(t)$ is the same as above.  The effect of the propagator is solely to induce a phase that varies as the first order Fourier coefficient, in addition to shifting all of the quasimomentum by an amount $\eta$.  In contrast, Eq. (\ref{eq:unitary_lin}) allows a broader class of unitaries to be synthesized.

\section{V. Summary and Outlook}

We have described the control of transport of atoms through a microwave-dressed spinor optical lattice.  Asymmetric lattices and spectral isolation provides a means to break the system up into a series of two-level systems, which aided the design of control routines.  We restricted our attention to translationally invariant systems, with no local addressing, but the possibility of a uniform applied time-dependent force. Under these conditions, we can determine the constraints of reachable states and more general reachable unitary maps that take a localized atom at one site to a state coherently extended over $n$ sites.    Based on these constraints, we propose a constructive protocol for carrying out these control tasks through a sequence of $SU(2)$ rotations acting in the two-level subspaces.
	
An important consideration for practical implementation of our protocol is robustness of the control sequences to imperfections in the system.  Because single particle transport is driven by a series of $SU(2)$ maps, certain errors may be fixed by borrowing techniques from robust control of NMR systems\cite{li_control_2006,cummins_tackling_2003,vandersypen_nmr_2005}.  In particular, spatial variations or miscalibrations in the microwave field strength and real or fictitious (i.e. light-shift induced) magnetic fields can be corrected with such techniques.  On the other hand, the inevitable spatial inhomogeneities in the lattice potential can lead to spatial variations in the energy common to both of the levels, which causes an overall phase on the two-level systems that varies in space.  Because this phase error is not an $SU(2)$ map, it cannot be removed with the standard NMR composite pulse protocols nor their generalizations, and we must develop new methods to correct this in order to make our protocol robust to lattice inhomogeneity.  This will be a topic of future investigation.
	
There are a number of ways in which our method can be extended to more general control of the transport of individual atoms. In the current work we restricted our attention to the single band, tight-binding approximation, in uniform lattices with two spin levels.  More general protocols that do not restrict the bands can be used to study the control of coupled spin and spatial degrees of freedom in a broader context.   Moreover, it has recently been shown \cite{merkel_quantum_2008} that the entire hyperfine manifold of magnetic sublevels is controllable with applied rf and microwave waveforms, leaving open the possibility of combining the control of high dimensional spin and spatial degrees of freedom.   In addition, breaking the translational invariance with quadratic or higher order potentials should allow the controllability of the system to be significantly enhanced.  Other modifications of the spin-dependent potentials, such as the spatial variation in microwave transition frequency that arises in a strong magnetic field gradient, can in principle allow spectral addressing of individual two-level systems and extend the controllability of the system.

Finally, the techniques proposed here may also be extended to control the dynamics of many-body systems.  For instance, it might be possible to use the microwave drive to synthesize more arbitrary interactions between atoms than are dictated by the static Hamiltonian.  Once interactions are included in the model, many-body unitary maps can by built from maps acting on restricted subspaces, as we have done here for single particles.   Such tools can play an essential role for quantum simulations of many-body Hamiltonians, both for studies of equilibrium properties such as the many-body phase diagram and non-equilibrium phenomena such as Lodschmidt echos \cite{cucchietti_loschmidt_2006} and the dynamics of phase transitions.

\emph{Acknowledgments}. We thank Seth Merkel for helpful discussions.  This research was supported by NSF Grants No.
PHY-0555573 and No. PHY-0555673, and IARPA/NIST Grant No. 70NANB4H1096.


\bibliographystyle{mischuck}
\bibliography{ControlledTransport}

\end{document}